\documentclass[aps,prd,preprint,showpacs,preprintnumbers]{revtex4}
\usepackage{graphics}
\def \beq{\begin{equation}}
\def \eeq{\end{equation}}
\def \beqa{\begin{eqnarray}}
\def \eeqa{\end{eqnarray}}
\def \det{{\rm Det}\,}
\def \tr{{\rm Tr}\,}
\def \ie{{\sl i.e.\/}}
\def \F{{\overline F}}
\def \Li{{\rm Li}}
\begin{document}

\title{Analyticity and the phase diagram of QCD}
\author{Sourendu \surname{Gupta}}
\email{sgupta@tifr.res.in}
\affiliation{Department of Theoretical Physics, Tata Institute of Fundamental
         Research,\\ Homi Bhabha Road, Mumbai 400005, India.}

\begin{abstract}
Some consequences of the analyticity of the free energy (pressure) of
QCD at finite chemical potential are deduced. These include a method
for numerical exploration of the full phase diagram by a novel use of
simulations at imaginary chemical potential to extract Yang-Lee zeroes
of the grand-canonical partition function. We make use of, and comment
on, CPT symmetries, positivity of non-linear susceptibilities and the
finiteness of screening lengths. We also comment on the structure of
zeroes expected for the usual picture of the phases of QCD, following
a discussion of the physics of imaginary chemical potential.
\end{abstract}
\preprint{TIFR/TH/03-0?, hep-lat/0307007}
\maketitle


Motivated by the complex phase structure of QCD at finite chemical
potential, $\mu$, predicted by effective theories and perturbative
analyses \cite{alford}, there has been a spate of work extending lattice
computations to finite $\mu$ \cite{fodor,owe,maria,allton,suscp}. Most of
these methods extrapolate data obtained at $\mu=0$ out to finite $\mu$
using, implicitly or explicitly, a Taylor expansion in $\mu$ of the
free energy. As a result, these expansions cannot be continued beyond
the nearest phase transition to $\mu=0$. Much of the interesting phase
structure of QCD then lies beyond their scope.  This paper explores
consequences of the analyticity of the free energy (guaranteed by
time reversal invariance) as a function of complex $\mu$ to suggest a
numerical technique for exploring the QCD phase diagram.  It turns out
that analyticity also has many other consequences, some of which are
touched upon.

The plan of this paper is the following. First it is shown how analyticity
is guaranteed by time reversal symmetry. This material is known to
practitioners, but serves to introduce notation and several concepts I
use later. Next I write down the canonical product representation for
the partition function and make contact with the Yang-Lee theory of
phase transitions. A numerical technique is developed to explore the
phase diagram using simulations at imaginary chemical potential, which
goes beyond using Taylor series for analytic continuation. The need for
a different approach is discussed next, with remarks on the phase diagram.


The partition function of QCD, \ie, $SU(N_c)$ gauge theory, with $N_f$
flavours of fermions, each subjected to a common chemical potential, $\mu$, is
\beq
   Z(T,\mu) = \int DU {\rm e}^{-S_g(T)}\prod_{f=1}^{N_f}\det M(m_f,\mu,T),
\label{partition}\eeq
where $M$ is the Dirac operator, $S_g$ the gauge part of the action and
the temperature $T$ enters the action through boundary conditions on
the fields. The free energy, $F$ and the pressure are defined by
\beq
   P(T,\mu) = -\,\frac1V F(T,\mu) = \left(\frac TV\right)\log Z(T,\mu).
\label{defs}\eeq

At any purely real or imaginary value of the chemical potential, the
free energy (or the pressure) is purely real \cite{weiss}. Is it possible
then for the free energy to be an analytic function of a general complex
$\mu$? Recall that this requires the real and imaginary parts of $F$,
$F_r$ and $F_i$ respectively, to satisfy the Cauchy-Riemann equations,
implying that they are conjugate harmonic functions. This rules out
$F_i=0$ unless $F$ is constant.

The answer is that it is entirely possible to have a harmonic function
(\ie, a function satisfying Laplace's equation) which is zero along
the real and imaginary axes and is non-trivial elsewhere.  Since all
even powers of the complex number $\mu$ satisfy these requirements,
it seems that analyticity of $F(\mu)$ (at fixed $T$) requires that it
be even in $\mu$.

Now recall that the transformation $\mu\to-\mu$ can be
compensated by time-reversal, \ie, by interchanging particles and
anti-particles. However, this definition is arbitrary; in the absence
of CP violating terms in the action, thermodynamics remains unchanged
by such a relabelling. The physics is therefore unchanged by changing
the sign of $\mu$. In other words, CP symmetry (same as time reversal,
by the CPT theorem) implies that $F(\mu)$ is an even function of $\mu$,
\ie, only even powers of $\mu$ appear in the Taylor expansion. Since the
whole chain of logic above can be inverted, this means that CP symmetry
is necessary and sufficient for the analyticity of $F$, and its reality
along the coordinate axes of $\mu$ in the complex plane.

The simplest use of analyticity is to expand the pressure in a Taylor
series \cite{allton,suscp} and to determine the coefficients by direct
lattice simulations at $\mu=0$ \cite{suscp,valence} or by fitting to
results obtained at imaginary $\mu$ \cite{maria}. However, this approach
is limited, since it only gives information on the phase connected to
$\mu=0$.  The first phase transition encountered stops the extrapolation,
and more machinery is required to get around this barrier. This is what
we develop in this paper.

At real $\mu$, the path integral measure in eq.\ (\ref{partition})
is complex \cite{wilczek} and importance sampling in a Monte Carlo
procedure fails although the integral is real. From the argument above,
it seems that by grouping together configurations related by CP, it
should be possible to construct a real weight by summing over each such
``CP orbit''.  Since CP symmetry forms a $Z_2$ group, each orbit consists
of exactly two configurations, and the sum of the weights in these two
configurations (for real $\mu$) must be real. The simplest example is
the Gibbs model \cite{gibbs}.

This model contains Fermions on a 1-dimensional lattice ($N_t$ sites in
the temporal direction) and $U(1)$ gauge fields on the links. Since there
are no plaquettes, the gauge action is trivial and the path integral can
be performed easily.  The model is a zero (spatial) dimensional field theory,
and, therefore is the quantum mechanics of a single fermion subjected to a
constant $U(1)$ field. It is possible to choose a gauge where all links
have equal value $U=\exp(i\theta)$. With fermion mass $m$, chemical
potential $\mu$ and defining $q^{N_t}=z=\exp(\mu/T)$, the Fermion matrix is---
\beq
   M=\left(\matrix{ m & qU & & \cdots & -U^\dag/q\cr
                    -U^\dag/q & m & qU & &\cdots \cr
                    & -U^\dag/q & m & qU & \cdots\cr
                    \vdots & & & \vdots \cr
                    qU & \cdots & & -U^\dag/q & m }\right).
\label{matrix}\eeq
The determinant is---
\beq
   \det M = w^{N_t} + 1/w^{N_t} - 2 \cosh\frac12 Nm'
\label{det}\eeq
where $w=qU$ and $ma/2=\sinh m'a/2$. The lattice spacing is $a=1/TN_t$.
The partition function is
\beq
   Z(N,z) = \int_{-\pi}^{\pi}\frac{d\theta}{2\pi}\det M,
\label{partgibbs}\eeq
which is real, although the weight is complex. In this simple model a
CP transformation induces the mapping $U\to U^\dag$. Then, summing over
each CP orbit one gets
\beq
   \sum_{CP}\det M = 4\cosh(\mu/T)\cos(N_t\theta) -4\cosh(m'/2T),
\label{cpsum}\eeq
which is real. In the partition function one should compensate the
double counting by dividing by a factor of two. This is immaterial
for expectation values.

In a more realistic model, the gauge configurations connected by a
CP transformation are harder to construct. 
It turns out to be easier to implement the symmetry
transformation C (which is the same as PT) which is local and maps
each link matrix $U\to U^\dag$. The T part of the transformation can be
used as before to prove that the free energy is even in $\mu$. Moreover,
this symmetry transformation preserves the values of $S_g$ and $\det M$
(see eq.\ \ref{partition}) separately. As a result, for the
Taylor expansion of $\det M$ summed over each PT orbit---
\beq
   \det M(m,T,\mu) = \det M(m,T,0) \left[ 1 + a_1\mu + \frac{a_2}2\mu^2
       +\frac{a_3}{3!}\mu^3 + \cdots\right],
\label{taylor}\eeq
one has $\Re\,a_i=\Im\,a_i=0$ for all odd $i$. In other words,
summed over PT orbits, the measure is even in $\mu$.

Since this summation over a symmetry orbit leads to a weight that is
the sum of two determinants, it is clear that the measure cannot be
rewritten using local pseudo-fermion fields.  Disappointingly, these
symmetries turn out to be unexploitable in this form for molecular
dynamics algorithms paralleling those which are used at zero chemical
potential \cite{algo}.  However, by ensuring analyticity, they lead to
other numerical approaches to the problem, as I show next.


From the Hamiltonian expression for the partition function, 
\beq
   Z(T,\mu)= \tr\exp\left[-\frac{\hat H}T-\frac{\mu\hat N}T\right],
\label{ham}\eeq
(where $\hat H$ is the Hamiltonian and $\hat N$ the number operator)
it is clear
that $Z$ has a periodicity of $2\pi T$ in the imaginary part of the
chemical potential---
\beq
   Z(T,\mu) = Z(T,\mu+2\pi ikT)\quad{\rm for\ any\ integer\ }k.
\label{period}\eeq
Functions with such periodicity are best analysed in terms of the
fugacity, $z=\exp(\mu/T)$. Entire functions in $\mu$ are regular
in the complex $z$ plane without the origin, \ie, the punctured $z$
plane \cite{notea}.  Then the partition function can be written in the
canonical product form \cite{ahlfors}
\beq
   Z(T,z) = {\rm e}^{-\F(T,z)/T}\prod\left(1-\frac z{z_n(T)}\right),
\label{zeroes}\eeq
where $z_n(T)$ are the zeroes of $Z$, and $\F$ is the regular part of
the free energy. CP symmetry constrains the set of zeroes, $\{z_n\}$
to be symmetric under inversions in the unit circle.  Branch points in
the full free energy, $F=-T\log Z$, then do not appear in the regular
part $\F$.  The periodicity of $Z$ in $\mu$ also implies that
\beq
   \F(T,\mu) = \F(T,\mu+2\pi ikT)\quad{\rm for\ any\ integer\ }k.
\label{phase}\eeq
The appropriate setting for further analysis is the Yang-Lee theory of
phase transitions \cite{yang}.

According to this theory, phase transitions occur at points in the
parameter space where dense sets of zeroes develop into pinch points.
In the present understanding of the phase diagram of QCD, one expects
pinch points on the real $\mu$ axis at small $T$ where interesting new
phases of QCD appear \cite{alford}.  We shall come back to this after
a digression on ideal gases which throws more light on the question of
periodicity in the imaginary part of $\mu$.

An ideal gas has the free energy---
\beq
   F(T,\mu) = N_cN_fV\left[ \frac{7\pi^2}{180} T^4 + \frac16 \mu^2T^2 
     + \frac1{12\pi^2} \mu^4\right],
\label{ideal}\eeq
which does not have the required periodicity; no polynomial in $\mu$
can. By examining the Hamiltonian formulation of an ideal gas, we can
show where the periodicity is hidden. The free energy can be written as
\beq
   F(\mu,T) = \frac{N_cN_fVT^4}{3\pi^2} \int_0^\infty dx x^3
     \left[\frac1{z{\rm e}^x+1}+\frac1{{\rm e}^x/z+1}\right].
\label{idealz}\eeq
Since $z$ is unchanged under $\mu\to\mu+2\pi iT$, this has explicitly
the required periodicity. By the substitution $y=\exp x$, the integral
can be written in terms of fourth order Euler Polylogarithm functions
\cite{li}---
\beq
   F(\mu,T) = -\frac{2N_cN_fVT^4}{\pi^2} \left[
        \Li_4\left(-\frac1z\right)+\Li_4\left(-z\right) \right],
\label{polylog}\eeq
which also explicitly retains the periodicity. In fact, each of the
terms has this periodicity explicitly. This property is lost
in a Taylor expansion of the polylogs in $\mu/T$---
\beq
   \Li_4\left(-{\rm e}^x\right) = -\frac{7\pi^4}{720} - \frac{3\zeta_3}4x -
       \frac{\pi^2}{24}x^2 - \frac{\log2}6x^3 - \frac{x^4}{48} - \frac{x^5}{480}
      +\frac{x^7}{40320}-\frac{x^9}{1451520}\cdots
\label{polyfour}\eeq
where the remaining terms are odd in the expansion parameter. The reason
is that in taking the logarithm of the argument one makes a choice of
the Riemann sheet, which is precisely equivalent to working within a
single strip of width $2\pi iT$.  When the sum in eq.\ (\ref{polylog})
is expanded in a Taylor series in $\mu$, it is clear that every term
cancels, apart from those which appear in eq.\ (\ref{ideal}). Periodicity
can be restored by averaging this expression over all Riemann sheets
of the logarithm, the result of which is expressed concisely in eq.\
(\ref{polylog}) \cite{noteb}. In general, interactions modify the integral
expression in eq.\ (\ref{idealz}), without spoiling the periodicity.
This upsets the conspiracy which cancels all higher terms in $F$,
leading to non-vanishing values for general Taylor coefficients, \ie,
the non-linear number susceptibilities \cite{suscp}.

We return now to considerations of phase transitions in QCD through
the Yang-Lee mechanism.  The set of zeroes, $z_n(T)$, of the partition
function contains the full information needed to identify all the phase
transitions in the theory through the Yang-Lee mechanism \cite{yang}. But
for all this to make sense, one needs to check that a thermodynamic
limit exists, and that, in this limit, the pressure is a continuous
and monotonically increasing function of $z$. This is Theorem I of
\cite{yang}, and is proven there in the non-relativistic limit when
interparticle potentials obey two conditions--- first that there should
be a limit to the number of particles that can be accommodated in a
finite box, and second that the interactions between them be of finite
range. Non-intuitive phenomena can occur when these conditions are
violated \cite{mukamel}. All work on QCD assumes the validity this
theorem. However, it is useful to first examine the basis for this
assumption.

In the low temperature phase, where the carriers of baryon number are
the baryons themselves, the problem is essentially non-relativistic,
since the baryon mass is much greater than the temperature. There is
ample experimental evidence that the inter-nucleon potential satisfies
both conditions necessary for the theorem to hold--- the baryon-baryon
interaction has a hard core repulsion and a short ranged Yukawa
interaction at long distances. In the high-temperature phase the problem
is relativistic, since one expects that the carriers of baryon number
are the quarks, whose masses are much less than the temperature. Quarks
and antiquarks are freely created and destroyed, and, as is well-known,
the correct reformulation of the problem is to work in ensembles where
the excess (or deficit) of quarks over antiquarks is fixed (the canonical
ensemble) or is conjugate to the chemical potential (the grand canonical
ensemble).  The second requirement, of finite range of interactions,
is straightforward. The gauge interaction, which could be long-ranged,
is known to be screened in the electric sector \cite{electric}. In the
magnetic sector there is no evidence for screening, but there is evidence
of confinement and hence a mass-gap \cite{magnetic}. All other screening
lengths are smaller in the continuum limit \cite{valence}. While these
phenomena are best established for $N_c=2$ and 3, they are expected to
hold also for other $N_c$.

If we accept these arguments, or otherwise directly assume the
first theorem of \cite{yang}--- that the pressure is a continuous
increasing function for $z\ge1$, then an immediate consequence is
that the Taylor expansion coefficients of $P(T,z)$ in $z$ at fixed $T$
are non-negative.  Now, the generalised susceptibilities \cite{suscp},
are Taylor coefficients of $P(T,\mu)$ in an expansion in $\mu$. Since
the expansion in $\mu$ can be obtained from that in $z=\exp(\mu/T)$
by expanding out the exponential, which is itself a convex function,
it follows that the sum of all generalised susceptibilities of a given
order are strictly positive.

This theorem of Yang and Lee is expected to hold for a continuum theory
in the thermodynamic limit of infinite volume. It is also expected to
hold for a continuum theory in a finite volume much bigger than the
range of interactions, even when the limit is taken with a sequence of
cutoffs, holding the physical volume fixed. On the other hand, at any
intermediate step in this procedure, the convexity argument may fail if
some of the screening masses are far from their continuum limits and not
sufficiently small.  Negative values of some of the susceptibilities
may then be observed at finite volume. However, convexity is regained
either by increasing the spatial volume or on taking the continuum
limit \cite{billoire}.

Further practical use of this theory necessitates estimation of the set
of zeroes, $\{z_n(T)\}$. Further information hinges on
the connection between the grand-canonical and
the canonical partition functions.  The partition function in eq.\
(\ref{partition}) generates expectation values in a grand canonical
ensemble with respect to quark number. A partition function for fixed
quark number, $Q$, \ie, in a canonical ensemble has also been used
before \cite{weiss,miller,barbour,hasenfratz}. One defines this canonical
partition function by the Fourier transform
\beq
   {\cal Q}(T,Q) = \frac1{2\pi}\int_{-\pi}^{\pi} d\phi Z(T,i\phi)\cos Q\phi,
\label{canon}\eeq
where $Z(T,i\phi)$ denotes the grand canonical partition function
in an imaginary chemical potential $\mu=i\phi$. One can invert this
formula to write
\beq
   Z(T,z) = \sum_{Q=1}^{VN_sN_fN_c} {\cal Q}(T,Q)\left[ z^Q+\frac1{z^Q}\right],
\label{grandc}\eeq
where $N_s$ is the number of components of the Dirac spinor ($N_s=4$ in
four dimensions). Note that on any lattice with $V$ sites on each spatial
slice, $Z$ is a polynomial in $z$ of degree $VN_sN_fN_c$.  The $Z_{N_c}$
symmetry of the gauge action can be used to show that ${\cal Q}_n=0$
unless $n$ is a multiple of $N_c$ \cite{hasenfratz}.  The remaining
coefficients are real and non-negative.  Hence the $VN_sN_f$ distinct
zeroes of $Z$ cannot lie on the positive real axis of $z$.  The Yang-Lee
picture now follows, with phase transitions occurring when $V\to\infty$
and the zeroes pinch the positive real $z$ axis. Note that the finite
Laurent series on the right of eq.\ (\ref{grandc}) is typical of a
relativistic theory with its CP symmetry. In non-relativistic theories
this is just a finite polynomial.

This gives a method for determining the transition line in the $(\mu,T)$
plane. Construct the partition function at imaginary chemical potential
and, by Fourier transforming it, build up the coefficients of the
polynomial in $z$. Transform this information into that of the zeroes of
the polynomial and check whether they begin to pinch the positive real $z$
axis. Since this information finds all the pinch points, this is one way
to go beyond the present day lattice computations and look for signals of
phase transitions to the conjectured colour superconducting phases. Some
results will be presented elsewhere.

This is an appropriate place to discuss a couple of interesting points
about this algorithm.  An absolute normalisation for $Z$ is not needed in
any practical implementation of this method.  An overall multiplicative
factor in $Z$ appears as such in the ${\cal Q}$'s, and hence is irrelevant
to the zeroes of eq.\ (\ref{grandc}). Simulations will never give an
absolute normalisation for $Z$.  However, combinations of simulations
at several different values of $\phi$ can be combined to give the
relative values of $Z$, as is done in setting up a multi-histogram
reweighting. There have been preliminary investigations of methods for
constructing $\cal Q$'s \cite{barbour,hasenfratz}.

A second practical question is that of statistical errors. Any simulation will
necessarily contain statistical errors, and the question naturally arises
whether they allow a determination of the pinch point of the Lee-Yang
zeroes with any degree of accuracy. Since the pinch points lie on the
real axis, after appropriate scaling, the zeroes in the vicinity of this point are
the solution of $z^N+1=0$, with $N$ being a large integer of the order
of $V$. The zeroes of this polynomial lie at the points $z_n=\exp[2\pi
i(n+1/2)/N]$.  The effect of simulation errors is to change this equation
to a general polynomial of order $k$---
\beq
   \sum_{i=0}^N \alpha_i z^i =0,
\label{genpo}\eeq
where the coefficients are real and we can write $|\alpha_N-1|
= \epsilon_N$, $|\alpha_0-1| = \epsilon_0$ and $|\alpha_i| =
\epsilon_i$ (for $0<i<N$). We model the errors $\epsilon_i$ as being
independent random numbers drawn from Gaussians of width $\sigma$. For
sufficiently large statistics $\sigma$ is small, and the deviations
of the roots of eq.\ (\ref{genpo}) from $z_n$, $\delta_n$, can be
treated as linear in the $\epsilon_i$. It is then a straightforward
exercise to show that $\langle\delta_n\rangle=0$ and the RMS error,
$\sqrt{\langle\delta_n^2\rangle-\langle\delta_n\rangle^2} = \sigma z_n/N$.
Thus the problem is statistically well conditioned. The estimates of
zeroes are unbiased and the errors fall inversely with the square root
of the statistics and inversely with the system volume.

We turn to the question of phase transitions at purely imaginary $\mu$,
since this yields further insight into the locations of the zeroes of the
partition function.  We note at the outset that crucial convexity theorems
(such as the second law) fail at imaginary $\mu$, and hence thermodynamics
in its usual sense does not apply. The idea of \cite{owe} is to use the
peak in a response function, $\chi$, (\ie, second derivative of the free
energy with respect to an intensive parameter) to identify putative
critical points at imaginary $\mu$ and finite volume, analytically
continue them to real $\mu$, and then take the infinite volume limit. Note
several subtleties in the argument. First, for a function $\chi(\mu^2)$,
the extrema for imaginary chemical potential, $\mu_*^2<0$, are obtained
by solving for the zeroes of the first derivative.  This does not
necessarily give the extrema at real chemical potential, since the
solution of $\chi'(\mu_*^2)=0$ does not guarantee that $\chi'(-\mu_*^2)$
also vanishes. A second, related subtlety can be seen in a double Taylor
expansion for any response function at finite volume---
\beq
   \chi(T/T_c,\mu/T) = \sum_{nm} c_{nm} \left(1-T/T_c\right)^n\,
       \left(\frac\mu T\right)^{2m},
\label{double}\eeq
where the Taylor coefficients $c_{nm}$ are independent of $T$ and
$\mu$ \cite{owe}. At the critical end-point, $T_E(\mu_E)$ this series
must sum to the divergent quantity $|1-T/T_E(\mu_E)|^{-\gamma}$
(with $\gamma>0$) in the infinite volume limit. There may also
be a critical point, $(T_E',\mu'_E)$ at an imaginary chemical
potential, $(\mu'_E)^2<0$. Clearly the sum in eq.\ (\ref{double})
differs at the two points, \ie, even if $\chi(T_E/T_c,\mu_E/T_E)$
and $\chi(T'_E/T_c,\mu'_E/T'_E)$ are both divergent, neither
$\chi(T_E/T_c,i\mu_E/T_E)$ nor $\chi(T'_E/T_c,\mu'_E/iT'_E)$ need
to diverge \cite{note1}.

\begin{figure}[hbt]\begin{center}
   \scalebox{1.0}{\includegraphics{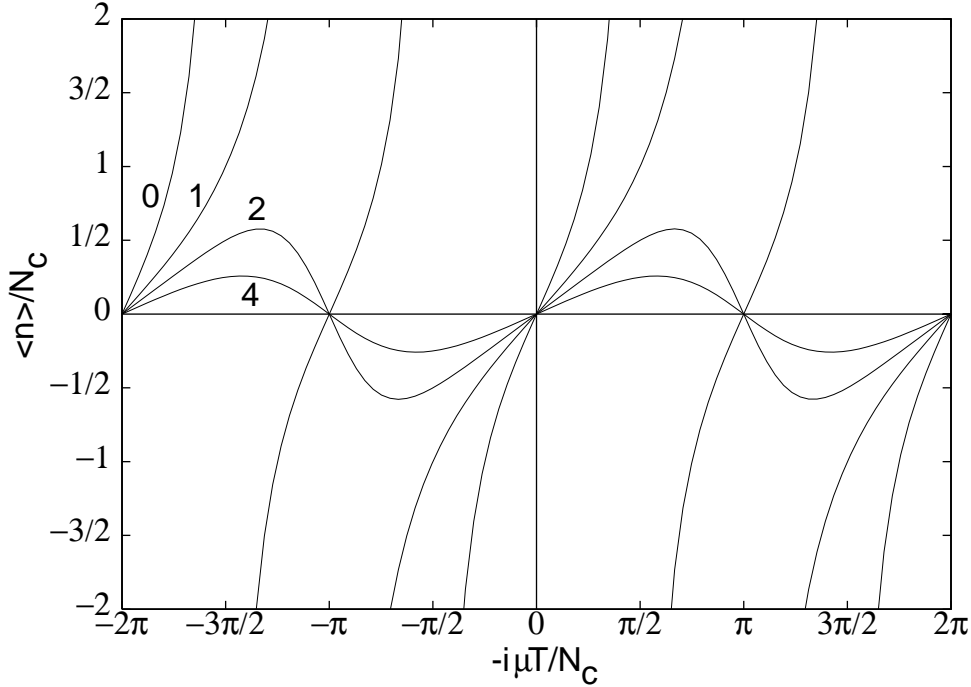}}
   \end{center}
   \caption{Number densities, $\langle n\rangle$ in the model of
      \cite{bilic} for the values of ${\cal Q}(T,0)$ marked in the
      figure.}
\label{fg.number}\end{figure}

In the context of our earlier arguments, it is more natural to look for
accumulation points of zeroes of the partition function for imaginary
chemical potential. This leads to the third subtlety, and a genuine
difference between real and imaginary chemical potential due to the lack
of convexity mentioned earlier. The theorem of Yang and Lee which we
have earlier quoted allowed us to express the partition function as a
finite (at all finite volumes) Laurent series in $z$ with non-negative
coefficients (eq.\ \ref{grandc}). This constraint on the coefficients
ensures convexity and prevents zeroes at real positive $z$, \ie, for real
$\mu$. Only in the limit of infinite volume can the zeroes pinch the real
line and cause a phase transition. However, this same structure allows,
(and, in some models, may force) the occurrence of zeroes on the unit
circle in $z$ (\ie, for imaginary $\mu$) even at finite volume.

Examples abound. We point out the extension of the Gibbs model to
$SU(N_c)$ gauge group where the partition function has been computed
\cite{bilic}, and found to be
\beq
   Z=2\cosh[N_c\mu/T] + \sinh[(N_c+1)m'/T]/\sinh[m'/T].
\label{bilic}\eeq
This is of the form in eq.\ (\ref{grandc}), with ${\cal Q}(T,N_c)=2$,
${\cal Q}(T,0)$ being the second term in the expression on the
right, and all other coefficients vanishing.  In the limit $m\to0$,
$Z=2\cosh[N_c\mu/T]$, and the zeroes of $Z$ are at
\beq
   z_n = {\rm e}^{i(2n+1)\pi/2N_c},\qquad(0\le n<N_c).
\label{bdzeroes}\eeq
Note that there are no positive real $z_n$. The zeroes lie exclusively on
the imaginary axis in the $\mu$-plane. The nearest zeroes to the positive
real $z$ axis are at $z=\exp(\pm i\pi/2N_c)$. In this 0+1 dimensional
problem there is no phase transition for any finite $N_c$. However,
a pinch point develops at $z=1$ as $N_c\to\infty$, very beautifully
illustrating van Hove's theorem.

\begin{figure}[tbh]\begin{center}
   \scalebox{0.5}{\includegraphics{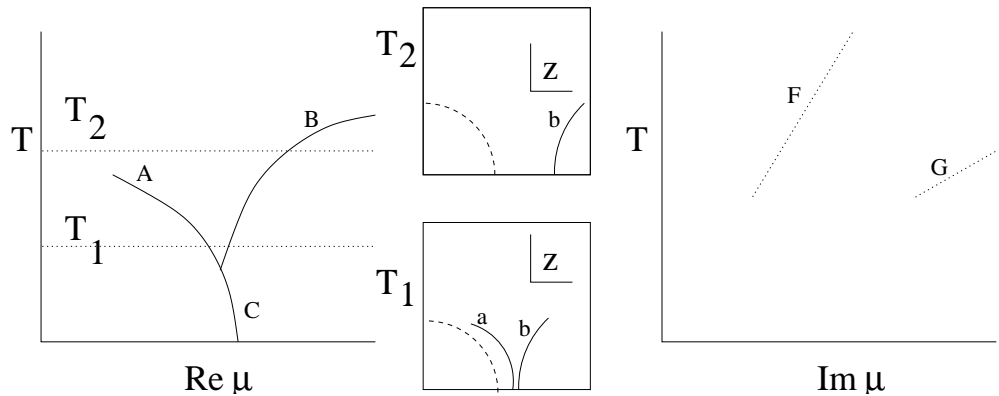}}
   \end{center}\vskip-5mm
   \caption{Phase diagrams along the real and imaginary axes in $\mu$
      for $N_c=3$ and $N_f=2$ for finite quark mass. The two insets show
      the lines of Yang-Lee zeroes pinching the phase transition points
      on the real $z$ axis. The phase diagram on the left constrains
      only the pinch points for the lines marked a and b, not their
      shapes and orientations. CP symmetry causes the whole phase diagram
      to be replicated for $\Re\mu<0$, and the corresponding Lee-Yang
      zeroes to be replicated inside the unit circle (dotted line) by
      inversion.}
\label{fg.phases}\end{figure}

The zeroes on the imaginary $\mu$ axis give rise to logarithmic branch
point singularities of the free energy evaluated at imaginary chemical
potential. The quark number density, all quark number susceptibilities,
the energy density and all its derivatives diverge at all these points for
all values of $N_c$.  These divergences are thermodynamically meaningless
since they occur without the necessity of taking a thermodynamic or
continuum limit. At finite quark mass, the zeroes remain on the unit
circle but move about as long as ${\cal Q}(T,0)\le1$. When the mass
becomes larger, these zeroes leave the imaginary axis, but they are still
visible as maxima of these observables (see Figure \ref{fg.number}),
and can be interpreted as crossover points. The effect of mass seems
to be one of the main differences between the perturbative and strong
coupling computations of \cite{weiss}.

These crossovers are generic at imaginary chemical potential--- Potts
models (or $O(N)$ models) with an ordering field in one direction
and an imaginary transverse field show such behaviour, as do massive
fermions subjected to imaginary chemical potential. Removing the first
ordering field converts them into branch point singularities. Both
pose dangers to numerical simulations, since these non-thermodynamical
singularities are easily misinterpreted as strong first order transitions,
or developing critical points.  If one indeed misidentifies such points,
then finite size scaling studies are also misleading, since one tries
to apply inappropriate homogeneity relations arising from thermodynamic
considerations. The idea of locating the zeroes of the partition function
using data collected at imaginary chemical potential is to bypass these
problems altogether.

The connection of the results of \cite{weiss} and the phase diagram of QCD
at real $\mu$ seems to be rather subtle. Here we make a hypothesis which
is consistent with the observed absence of a critical point with $Z_3$
symmetry \cite{z3} and with present understanding of the physics of the
$Z_{N_c}$ phases \cite{kiskis}.  Figure \ref{fg.phases} is a cartoon of
the conjectured phases of QCD ($N_c=3$ and $N_f=2$) with finite quark mass
\cite{alford}. There are three lines of first order phase transitions
(labelled A, B and C) in the figure. These meet at a triple point,
where the normal phase, the plasma phase and a colour-superconducting
phase coexist. The line A, separating the normal and plasma phases ends
in a critical point \cite{note2}.  The line $B$ is expected to continue
to infinity. For imaginary chemical potential, only the locus of zeroes
of the partition function, the lines F and G, are shown.

There are also two insets showing the positions of Yang-Lee zeroes which
might lead to such phase diagrams. In this figure only the pinch points
are constrained--- the remainder of the lines 'a' and 'b' are allowed to
twist, merge or intersect without changing the phase diagram \cite{borgs}.
The important point is that in a range of temperatures, there are two
pinch points on the real axis \cite{note3}.  As the temperature rises,
the pinch point on the right, labelled 'b', moves to higher and higher
$\mu_r$, whereas the one on the left, labelled 'a', lifts away from the
real axis at the critical end point, $T_E$.  Its effect can still be
felt at the crossover temperature, $T_c$, seen in finite temperature
simulations at $\mu=0$.  Above $T_c$ the line 'b' continues to move
outwards, but there remain isolated $N_c$ zeroes on or near the unit
circle (drawn with a dotted line), depending on the quark mass, which
give rise to the crossovers (F and G) observable at imaginary $\mu$.


We end by summarizing the contents of this paper.  The complex analytic
structure of the free energy (pressure) of QCD was investigated. It
was shown that CP symmetry makes the free energy an analytic function
of the chemical potential $\mu$ at all temperatures $T$. This symmetry
is unexploitable to give an algorithm for a Monte Carlo procedure that
works through a molecular dynamics algorithm. Exploiting analyticity, it
is possible to write a canonical product representation of the partition
function, and made contact with the Yang-Lee theory.  It was argued
that analyticity, monotonicity and continuity of the pressure implies
that non-linear susceptibilities of arbitrary order must exist and be
non-negative at all $T$. It was further argued that this is connected
with electric screening and magnetic confinement in high temperature QCD.
I showed how to use canonical partition functions, derived from Monte
Carlo simulations at purely imaginary chemical potential, to obtain
information on the phase structure of QCD. Finally, a plausible phase
diagram of QCD was used to anticipate the behaviour of the zeroes that
such numerical techniques might give rise to.


\begin{thebibliography}{99}
\bibitem{notea}
   CP symmetry then implies that $Z$ is symmetric under inversions
   in the unit circle, $z\to1/z$. In the remaining part of this paper all
   statements should be interpreted as pertaining to one part of the $z$
   plane unless otherwise stated. We will usually examine the region outside
   the unit circle, \ie, for $\Re\mu>0$. By inversion, \ie, CP
   transformations, the physics inside the unit circle can be reconstructed.
\bibitem{noteb}
   In a path-integral approach one obtains directly the expression in eq.\ 
   (\ref{ideal}) and periodicity can only be obtained then by summing
   over all the strips of width $2\pi iT$. The enhanced periodicity
   noticed in \cite{weiss} implies that one sums over strips of width
   $2\pi iT/N_c$.
\bibitem{note1}
   These are straightfoward consequences of comparison of general alternating
   series and absolutely convergent series.
\bibitem{note2}
   At zero quark mass the critical point may be pushed to $\mu=0$, or may change
   into a tricritical point from which a critical line emerges. While there are
   theoretical biases towards the latter alternative, deciding between them is
   one of the goals of future lattice simulations.
\bibitem{note3}
   For certain classes of models the zeroes lie only on a circle. This gives a
   single phase transition at any temperature. QCD is expected to fall outside
   this class of theories. However, at the triple point, the two pinch points
   are expected to merge. It is interesting to speculate that across the triple
   point the potential between effective baryonic degrees of freedom changes in
   character.
\end{thebibliography}
\end{document}